\documentclass[aps,prd,twocolumn,showpacs,eqsecnum,amsmath,nofootinbib]{revtex4}

\usepackage{graphicx,amssymb,amsthm,amsmath}

\def \d {{\rm d}}
\def \e {e}

\newcommand{\norm}[1]{\|#1\|}
\newcommand{\abs}[1]{|#1|}

\begin{document}

\title{Past horizons in D-dimensional Robinson--Trautman spacetimes}

\author{Otakar Sv\'{\i}tek}

\affiliation{Institute of Theoretical Physics, Faculty of Mathematics and Physics, Charles University in Prague, V~Hole\v{s}ovi\v{c}k\'ach 2, 180~00 Prague 8, Czech Republic }

\email{ota@matfyz.cz}

\begin{abstract}
We derive the higher dimensional generalization of Penrose--Tod equation describing past horizon in Robinson--Trautman spacetimes with a cosmological constant and pure radiation. Existence of its solutions in $D>4$ dimensions is proved using tools for nonlinear elliptic partial differential equations. We show that this horizon is naturally a trapping and a dynamical horizon. The findings generalize results from $D=4$.
\end{abstract}

\date{\today}

\maketitle

\section{Introduction}

Robinson--Trautman spacetimes represent a class of expanding nontwisting and nonshearing solutions \cite{RobinsonTrautman:1960,RobinsonTrautman:1962,Stephanietal:book} describing generalized black holes. Various aspects of this family in four dimensions have been studied in the last two decades. In particular, the existence, asymptotic behaviour and global structure of \emph{vacuum} Robinson--Trautman spacetimes of type~II with spherical topology were investigated, most recently in the works of Chru\'{s}ciel and Singleton \cite{Chru1,Chru2,ChruSin}. 
In these rigorous studies, which were based on the analysis 
of solutions to the nonlinear Robinson--Trautman equation for generic, arbitrarily strong smooth initial data, the spacetimes were shown to exist globally for all positive retarded times, and to converge asymptotically to a corresponding Schwarzschild metric. Interestingly, extension  across the  ``Schwarzschild-like'' future event horizon can only be made with a finite order of smoothness. 
Subsequently, these results were generalized in \cite{podbic95,podbic97} to
the Robinson--Trautman vacuum spacetimes which admit a nonvanishing 
 \emph{cosmological constant} $\Lambda$. These cosmological solutions settle down exponentially fast to a Schwarzschild--(anti-)de Sitter solution at large times $u$. 
Finally, the  Chru\'{s}ciel--Singleton 
analysis was extended to Robinson-Trautman spacetimes including matter, namely 
\emph{pure radiation} \cite{PodSvi:2005}. It was demonstrated that these solutions with pure radiation and a cosmological constant 
exist for any smooth initial data, and that they approach the spherically symmetric Vaidya--{(anti-)}de~Sitter metric. 

In \cite{podolsky-ortaggio}, Robinson--Trautman spacetimes (containing aligned pure radiation and a cosmological constant $\Lambda$) were generalized to any dimension. The evolution is governed by a simpler equation in higher dimensions, contrary to the four-dimensional case where fourth order parabolic type Robinson--Trautman equation occurs. Also, the possible algebraic types were determined. But still several interesting features deserve attention, the presence of horizons being among them. Similarly to four dimensions, higher-dimensional Robinson--Trautman family of solutions contains several important special cases, e.g. Schwarzschild--Kottler--Tangherlini black holes and generalizations of the Vaidya metric. The study of higher-dimensional spacetimes and their features help to comprehend which properties survive the generalisation and which are closely tied to four dimensions, thus deepening the understanding of General Relativity. Lately, considerable interest in higher dimensions comes from outside of the purely relativistic community.

Our concern here is to locate the past (white hole) horizon. In general dynamical situations this might be rather nontrivial since the obvious candidate - event horizon - is a global characteristic and therefore the full spacetime evolution is necessary in order to localize it. Therefore, over the past years different quasilocal characterizations of black hole boundary were developed. The most important ones being apparent horizon \cite{hawking-ellis}, trapping horizon \cite{hayward} and isolated or dynamical horizon \cite{ashtekar,krishnan}. The basic {\it local} condition in the above mentioned horizon definitions is effectively the same: these horizons are sliced by marginally trapped hypersurfaces with vanishing expansion of outgoing (ingoing) null congruence orthogonal to the surface. Quasilocal horizons are frequently used in numerical relativity for locating the black holes or in black hole thermodynamics.

For the vacuum four dimensional Robinson--Trautman solutions without cosmological constant the location of the horizon together with its general existence and uniqueness has been studied by Tod \cite{tod}. Later, Chow and Lun \cite{chow-lun} analyzed some other useful properties of this horizon and made numerical study of both the horizon equation and Robinson--Trautman equation. These results were recently extended to nonvanishing cosmological constant \cite{PodSvi:2009}.

\section{Robinson--Trautman spacetime in D dimensions}

Robinson--Trautman spacetimes (containing aligned pure radiation and a cosmological constant $\Lambda$) in any dimension were obtained in \cite{podolsky-ortaggio} using the geometric conditions of the original articles about the four-dimensional version of the spacetime \cite{RobinsonTrautman:1960,RobinsonTrautman:1962}. Namely, they required the existence of a twistfree, shearfree and expanding null geodesic congruence. They have arrived at the following metric valid in higher dimensions
\begin{equation}
 \d s^2=\frac{r^2}{P^2}\,\gamma_{ij}\,\d x^i\d x^j-2\,\d u\d r-2H\,\d u^2 
\end{equation}
where $2H=\frac{{\cal R}}{(D-2)(D-3)}-2\,r(\ln P)_{,u}-\frac{2\Lambda}{(D-2)(D-1)}\,r^2-\frac{\mu(u)}{r^{D-3}}$. The unimodular spatial $(D-2)$-dimensional metric $\gamma_{ij}(x)$ and the function $P(x,u)$ must satisfy the field equation ${\cal R}_{ij}=\frac{{\cal R}}{D-2}h_{ij}$ (with $h_{ij}=P^{-2}\gamma_{ij}$ being the rescaled metric) and $\mu(u)$ is a ``mass function'' (we assume $\mu>0$). In $D=4$ the field equation is always satisfied and ${\cal R}$ (Ricci scalar of the metric $h$) generally depends on $x^i$. However, in $D>4$ the dependence on $x^i$ is ruled out (${\cal R}={\cal R}(u)$). But generally, it still allows a huge variety of possible spatial metrics $h_{ij}$ (e.g., for ${\cal R} > 0$ and $5 \leq D-2 \leq 9$ an infinite number of compact Einstein spaces were classified). The dynamics is also different in $D>4$. While in four dimensions there is a fourth order Robinson--Trautman equation, the corresponding evolution equation is much easier in higher dimensions
\begin{equation}                     
(D-1)\,\mu\,(\ln P )_{,u}-\mu_{,u} =\frac{16\pi n^{2}}{D-2}\ ,
\end{equation}
where function $n$ describes the aligned pure radiation. 

\section{Past horizon}

\begin{figure}[ht]
\centering
\includegraphics[scale=0.8]{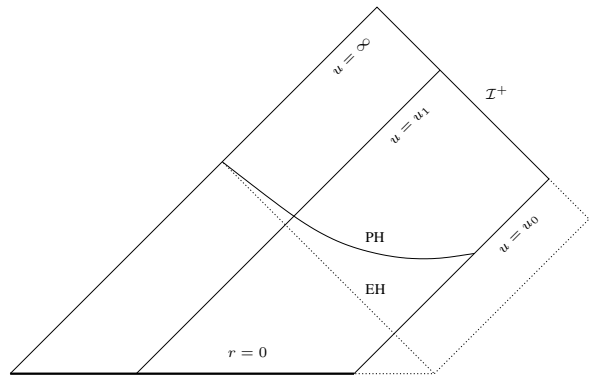} 
\caption{Schematic conformal diagram of Robinson-Trautman spacetime in $D=4$ with $\Lambda=0$ and indicated past (trapping) horizon (PH) and event horizon (EH).}\label{fig}
\end{figure}

In our case, we will be dealing only with the condition of vanishing expansion defining the marginally trapped hypersurfaces. Concretely, we will search for the past horizon similarly to previous studies in four dimensions and corresponding to the form of the metric containing retarded time. As will be clear later, we might call it {\it trapping horizon} or even {\it dynamical horizon} if it is spacelike, assuming appropriate higher-dimensional generalization of these notions (see \cite{Cao}). In four-dimensional case the parabolic character of Robinson--Trautman equation makes it generally impossible to extend the spacetime to past null infinity (the solutions of the Robinson--Trautman equation are generally diverging when approaching $u=-\infty$) and it is impossible to define event horizon. In higher dimensions this is no longer truth (the evolution equation is different) but since one would like to investigate the horizon existence generically, without prior specification of all necessary functions (e.g. dynamics of pure radiation) and geometry of $(D-2)$-dimensional spatial hypersurfaces, the best approach is still using the quasilocal horizons.  In figure \ref{fig}, the schematic conformal picture of Robinson--Trautman spacetime (for $D=4$ and without cosmological constant for simplicity) is presented together with the approximate location of the horizons (initial data are given at $u=u_{0}$).

The explicit parametrization of the {\em past horizon} hypersurface is $r=R(u,x^{i})$ such that its intersection with each $u=u_{1}$ slice is an outer marginally past trapped ($D-2$)-surface.

For the calculation of the expansion of an appropriate null congruence we will use a straight-forward generalization of the tetrad formalism to arbitrary dimension. Note that one can no longer use complex vector notation. Using two null covectors $l_{a}, n_{a}$ (with normalization $l_{a} n^{a}=-1$) and $D-2$ spatial covectors $m_{a\{i\}}$ ($i=1,..,D-2$) we suppose the following decomposition of the metric
\begin{equation}
g_{ab}=-2\,l_{(a}n_{b)}+m_{a\{i\}}m_{b\{j\}}\,\delta^{ij}
\end{equation}
Null D-ad (D-bein) adapted to the trapped hypersurface (using the above mentioned parametrization) has the following form:
\begin{eqnarray}
l^{a}&=&(0,1,0,..,0)\nonumber\\
n^{a}&=&\left(1,\,[-H+{\textstyle \frac{r^2}{2}}\, h({\mathbf \nabla}R,{\mathbf \nabla}R)]\, ,{\mathbf \nabla} R\right)\label{D-ad}\\
m^{a}_{\{i\}}&=&\left(0,\,Pr\, h({\mathbf \nabla}R,{\mathbf w}_{i})\,,{\textstyle \frac{1}{r}}{\mathbf w}_{i}\right)\nonumber
\end{eqnarray}
where $D-2$ vectors ${\bf w}_{i}$ diagonalize metric $h$, ${\bf \nabla} R=\{R^{,x^{1}},..,R^{,x^{D-2}}\}$ and $h(\cdot ,\! \cdot)$ denotes scalar product w.r.t.$\;h$. Fortunately, in subsequent calculations we do not need the explicit form of the vectors ${\bf w}_{i}$, it is sufficient to know their orthogonality properties.

By straight-forward computation one easily calculates the expansion associated with the congruence generated by $l^{a}$ to be $\Theta_{l}= \frac{D-2}{r}$ meaning that the outgoing null congruence is diverging. This is exactly what one assumes when dealing with the past trapped surface and is the additional condition in the definition of trapping horizon \cite{hayward}.

\section{Generalized Penrose--Tod equation}
Ingoing null congruence expansion can be calculated using the formula (sometimes a $(D-2)$ factor is used in the definition, but we are going to evaluate it to zero anyway) $\Theta_{n}=n_{a;b}\,p^{ab}$, where the tensor $p^{ab}=g^{ab}+2\,l^{(a}n^{b)}$ corresponds to the hypersurface projector. From $\Theta_{n}= 0$ (equivalent to Penrose--Tod equation in four dimensions) we get the marginally trapped hypersurface condition
\[
{\cal R}-{\textstyle \frac{2(D-3)}{D-1}}\Lambda R^{2}-{\scriptstyle (D-2)(D-3)}\frac{\mu}{R^{D-3}}{\scriptstyle -{2(D-3)}}\Delta(\ln R)-
\]
\begin{equation}\label{PT}
-{\scriptstyle (D-4)(D-3)}\,h(\nabla \ln R,\nabla \ln R)= 0
\end{equation}
It is a nonlinear second order partial differential equation, where both the laplacian and scalar product in the last term correspond to the Einstein metric $h_{ij}$. Interesting property of this equation is that for $D>4$ its nonlinearity is much worse since the term quadratic in derivatives appears.

\subsection{Results for $D=4$}
In four-dimensional case one can no longer use the existence proof for equation (\ref{PT}) given by Tod \cite{tod} when the cosmological constant is present. However, one can use the version of sub and super-solution method adapted to Riemannian manifolds by Isenberg \cite{isenberg} and valid for equations of the form $\Delta \psi=f(x,\psi)$. For the proof of uniqueness one may use a straightforward modification of the original Tod's proof \cite{tod} incorporating the cosmological constant. Using Newmann-Penrose equations one can also determine the character of the horizon as a three-dimensional hypersurface. These results (for $\mu=2m=const.$) are derived in \cite{PodSvi:2009} and summarized in the following table:

\begin{table}[h]
\caption{$D=4$}\label{table}
$$
\begin{array}{||c||c|c|c||}
\hline \hline
\mbox{RESULTS} & \Lambda=0 & \Lambda<0 & \Lambda>0 \\
\hline \hline
\mbox{Existence} & \mbox{Always} & \mbox{Always} & \Lambda < \frac{4}{9\mu^2} \\
\hline
\mbox{Uniqueness} &  \mbox{Always} & \mbox{Always}  & R < \sqrt[3]{\frac{3\mu}{2\Lambda}}\\
\hline
\mbox{Spacelike or null} &  \mbox{Always} & \mbox{Always}  & R < \sqrt[3]{\frac{3\mu}{2\Lambda}}\\
\hline \hline
\end{array}
$$
\end{table}

The restrictions for the positive cosmological constant can be easily understood by specializing to spherical symmetry and $\Lambda>0$ (Schwarzschild de--Sitter) :
\begin{itemize}
\item $\Lambda < \frac{4}{9\mu^2}=\frac{1}{9m^2}$ rules out an over-extreme case.
\item $R < \sqrt[3]{\frac{3\mu}{2\Lambda}}=\sqrt[3]{\frac{3m}{\Lambda}}$ for the extreme case ($9\Lambda m^2=1$) reduces to $R < 3m$ which may be interpreted as showing the uniqueness of the past black/white hole horizon (as opposed to the cosmological one).
\end{itemize}
Both explanations are quite natural and not surprising.

\subsection{$D>4$ : Existence of the solution}

The methods used in $D=4$ are not applicable when the equation is of the form (after the substitution $R=C e^{-u}$ in (\ref{PT}), assuming $u\geq 0$ with a suitable constant $C$)
\begin{equation}
\Delta u=F(x,u,\nabla u)\ ,
\end{equation}
where $F$ is quadratic in gradient.

To prove existence of the solution to this quasilinear equation we will proceed by combining several steps (motivated by \cite{Kuo} and using results from \cite{Besse,Boccardo,Gilbarg-Trudinger}).

\begin{itemize}
\item[1.] We will consider the differential operator $P u=-2(D-3)\Delta u+\rho u$, with $\rho>0$ on a Riemannian manifold $M$. By using Maximum Principle we can prove that $\ker(P)=0$ \cite{Besse}.\newline
\item[2.] The linear differential equation $P u=f$ with $f\in C^{0,\alpha}(M)$ (H\"{o}lder space over $M$) has unique solution  $u\in C^{2,\alpha}(M)$ (this standard result can be proven for example by Fredholm alternative\cite{Besse} and the previous step). \newline
\item[3.] To proceed with the nonlinear problem $P u=f(x,u,\nabla u)$, with $f$ determined from (\ref{PT}) as ($\norm{\cdot}_{h}$ stands for the norm with respect to the positive definite metric $h_{ij}$)
\[
f=-\rho u + {\cal R}-{\textstyle \frac{2(D-3)}{D-1}}\Lambda C^{2}\e^{-2u}-
\]
\[
\qquad -{\scriptstyle (D-2)(D-3)}{\mu}{C^{3-D}}\e^{(D-3)u}-{\scriptstyle (D-4)(D-3)}\norm{\nabla u}_{h}^2\ ,
\]
 we introduce the following truncature \cite{Kuo,Boccardo} :\newline
$f_{n}$ - truncature of $f$ by $\pm n$.\newline
Then the map $v\in C^{1,\beta}(M)\to f_{n}(x,v,\nabla v)$ is bounded. Using the previous step together with results on composition of H\"{o}lder functions there exists a unique $w\in C^{2,\alpha\beta}(M)$ solving $P w=f_{n}(x,v,\nabla v)$.\newline
\item[4.] The map $v\to w$ from previous step satisfies conditions of Schauder Fixed Point theorem, namely the {\em a priori} boundedness (see \cite{Besse} or \cite{Gilbarg-Trudinger}) $\Rightarrow$ for each $n$ there is a fixed point $u_{n}\in C^{1,\beta}(M)$ (even $u_{n}\in C^{2,\alpha\beta}(M)$) solving $P u_{n}=f_{n}(x,u_{n},\nabla u_{n})$ and moreover one can easily verify that $\norm{u_{n}}_{L^{\infty}}\leq \frac{n}{\rho}$ (considering $P u_{n}=f_{n}(x,v,\nabla v)\leq n$ and compactness for evaluation at the maximum of $u_{n}$).\newline
\item[5.] Using results of Boccardo, Murat \& Puel \cite{Boccardo}, in particular their Proposition 3.6, we can state the following corollary: \newline
Assuming that metric $h_{ij}$ is smooth, function $F$ can be estimated like $\abs{F} \leq B(u)(1+\abs{\nabla u}^2)$ (where $B(u)$ is increasing function on ${\mathbb R}^{+}$), and there exist a sub- and a super-solution \cite{footnote} $u^{-}\leq u^{+}$, $u^{\pm}\in C^{1,\beta}(M)\cap L^{\infty}(M)$, then there is a $L^{\infty}$-bounded subsequence $u_{\bar{n}}$ of the approximating solutions from the previous step satisfying $u^{-}\leq u_{\bar{n}}\leq u^{+}$ a.e.\newline
Indeed, inspecting the above defined function $f=\rho u-2(D-3)F$ one can verify that function $B(u)$ might be found, namely there is no singular behaviour at $u=0$. Also, the domain we are dealing with is compact and therefore any dependence on $x$ can be bounded for well behaved objects we use. For example, one may select the following bounding function
\[
B(u)=\max\left(W,{\textstyle \frac{D-4}{2}} \max_{x\in M}\norm{h_{ij}(x)}\right)+
\]
\[
+{\textstyle \frac{(D-2)\mu C^{3-D}}{2}} \e^{(D-3)u}\ ,
\]
where $W=\left|{\cal R}-{\textstyle \frac{\Lambda C^2}{D-1}}-{\textstyle \frac{(D-2)\mu C^{3-D}}{2}}\right|$ and the matrix norm of $h$ was used.
\newline
\item[6.] Thanks to elliptic estimate $\norm{u_{\bar{n}}}_{C^{2,\gamma}}\leq K(\norm{u_{\bar{n}}}_{C^{0}}+\norm{f_{\bar{n}}}_{C^{0,\gamma}})\leq K(\norm{u_{+}}_{C^{0}}+N\norm{f}_{C^{0,\gamma}})$ it is even $C^{1,\beta}$-bounded. To estimate $f_{\bar{n}}$ in the last inequality one can use its representation as $f_{\bar{n}}=fg_{\bar{n}}$, where function 
\[
\qquad\ {\textstyle g_{\bar{n}}=1-\Theta(f-\bar{n})\left(1-\frac{\bar{n}}{f}\right)-\Theta(-f-\bar{n})\left(1+\frac{\bar{n}}{f}\right)}
\]
is responsible for the truncation and $\Theta$ is the Heaviside function. Using the results for composition (e.g. H\"{o}lder index of composed map is a product of indices of components) and multiplication (e.g. index is a minimum of indices of components) of H\"{o}lder continuous functions on bounded sets there has to be a new H\"{o}lder coefficient $\gamma\leq\alpha\beta$ and a suitable constant $N$ fulfilling the inequality.
Function $f$ in the elliptic estimate is dependent on $x$ not only explicitly but also via $u_{\bar{n}}(x)$ and $\nabla u_{\bar{n}}(x)$, which is reflected in the constant $N$. While $u_{\bar{n}}$ is bounded independently on $\bar{n}$ by $u^{\pm}$ we need to bound the gradient in the same way. Using the fact that our function $F$ has the form $F_{1}(u)+F_{2}(x)\norm{\nabla u}_{h}^2$ (with strictly positive $F_{2}$) and integrating over the manifold (using the Stokes theorem to eliminate the laplacian) we get
\[
-\int_{M}F_{1}(u_{\bar{n}})=\int_{M}F_{2}(x)\norm{\nabla u_{\bar{n}}}_{h}^2\geq
\]
\[
\geq F_{2,min}\int_{M}\norm{\nabla u_{\bar{n}}}_{h}^2\ ,
\]
where the left hand side might be independently estimated. According to previous results gradient of $u_{\bar{n}}$ is bounded and from the last equation even independently. Therefore, the constant $N$ does not depend on  ${\bar{n}}$.
\newline
\item[7.] Then there is a $C^{1,\beta}$-convergent subsequence $u_{\tilde{n}} \to u^{s}$, which proves the existence of the solution provided the sub- and super-solutions are obtained. Moreover, using the second step with $f(x,u^{s},\nabla u^{s})$ we must have  $u^{s}\in C^{2,\alpha\beta}(M)$.
\end{itemize}

As is most common in the literature, we would be looking for constant sub- and super-solutions, first in the case ${\cal R}>0$ (assuming $u_{min}\geq 0$ which can always be arranged by a suitable choice of $C$) :
\begin{itemize}
\item ${\Lambda\leq 0}$
\[{\textstyle u_{1}^{+}=\frac{1}{D-3}\ln\left[\frac{C^{D-3}}{(D-2)(D-3)\mu}{\cal R}\right]\ ,}\]
\[{\textstyle u_{1}^{-}=\frac{1}{D-3}\ln\left[\frac{C^{D-3}}{(D-2)(D-3)\mu}({\cal R}-\frac{2(D-3)}{D-1}\Lambda C^2)\right]\ ,}\]
\item ${\Lambda > 0}$
\[u_{2}^{+}=u_{1}^{-}\ ,\]
\[u_{2}^{-}=u_{1}^{+}\ .\]
\end{itemize}
These solutions satisfy all the conditions for any $\Lambda\leq 0$, but for positive cosmological constant one has to demand 
\begin{equation}\label{restriction}
{\textstyle {\frac{2 {\cal R}}{(D-1)(D-2)(D-3)\mu}}\left(\frac{{\cal R}}{2 \Lambda}\right)^{\frac{D-3}{2}}\geq 1\ },
\end{equation}
so that $u_{2}^{+}\geq 0$ is valid for such a constant $C$ which is maximizing the value of $u_{2}^{+}$. Interestingly, this last condition reduces in the four-dimensional case (that was not explicitly studied here, but can be included trivially - note that then ${\cal R}$ is not a constant on $u=const.$ and $u^{\pm}$ has to be adjusted \cite{PodSvi:2009}) to the condition from the table \ref{table} for the existence of the solution when $\Lambda > 0$. One has to remember that for $D=4$ scalar curvature ${\cal R}$ asymptotically (as $u\to\infty$) approaches value $2$.

One may wonder whether the condition (\ref{restriction}) is necessary or if it might be weakened by the choice of more suitable non-constant sub- and/or super-solution. Let us assume we have a positive solution $R\in C^{2,\delta}(M)$ of (\ref{PT}) with $\Lambda > 0$. Since it represents a function on a compact manifold it has to attain its maximum $R_{max}$ and minimum $R_{min}$. At $R_{min}$ the gradient term in (\ref{PT}) vanishes while $\Delta (\ln R_{min})\geq 0$ leading to the following inequality
\begin{equation}\label{Rmin}
-{\cal R}+{\textstyle \frac{2(D-3)}{D-1}}\Lambda R_{min}^{2}+{\scriptstyle (D-2)(D-3)}\mu R_{min}^{3-D}\leq 0
\end{equation}
The right-hand side of (\ref{Rmin}) has minimum at $R_{min,E}^{D-1}=\frac{(D-1)(D-2)(D-3)\mu}{4\Lambda}$. The inequality (\ref{Rmin}) must also hold for this value and after its substitution one arrives exactly at the condition (\ref{restriction}). Therefore, it represents not only sufficient but also necessary condition for the existence of the horizon when positive cosmological constant is present. 

Since according to mathematical results any manifold (including compact ones) of dimension greater than or equal to $3$ can be endowed with a complete Riemannian metric of constant negative scalar curvature \cite{Aubin,Lohkamp} one should also consider that ${\cal R}<0$ for our $(D-2)$-dimensional spatial hypersurface. One can propose the following constant sub- and super-solutions for ${\cal R}\leq 0$ (assuming $u_{min}\geq 0$)

\begin{itemize}
\item ${\Lambda< 0}$
\[u_{3}^{+}=0\]
\[
{\textstyle u_{3}^{-}=\frac{1}{D-3}\ln\left[-\frac{2 C^{D-1}}{(D-1)(D-2)\mu}\Lambda\right]}
\]
and select $C\geq C_{min}$ that is defined by
$${\cal R}-{\textstyle \frac{2(D-3)}{D-1}}\Lambda C^{2}_{min}-{\scriptstyle (D-2)(D-3)}\mu C_{min}^{-(D-3)}=0\ ,$$

\item ${\Lambda \geq 0}$ : impossible to find constant $u^{+}$.
\end{itemize}
So for nonpositive scalar curvature ${\cal R}$ we can prove the existence only for negative cosmological constant.

In the four-dimensional case one can infer some useful results from the previous inequalities, mainly due to the fact that one can bound the scalar curvature from the asymptotic behaviour or using Gauss-Bonnet theorem. However, in higher-dimensional spacetime neither tool is available (the generalizations of Gauss-Bonnet theorem are very complicated and not immediately applicable).

\subsection{$D>4$ : Character of the horizon}
After establishing the existence of the past horizon $\mathcal{H}$ as a hypersurface foliated by marginally trapped surfaces one is naturally interested in whether it satisfies other conditions of recent quasilocal horizon definitions. We will consider trapping and dynamical horizons.

The previous results tell us that $\Theta_{l}>0$ and $\Theta_{n}=0$ holds on the past horizon. Since the Lie derivative $\mathcal{L}_{l}\Theta_{n}$ is in general nonvanishing on the horizon its closure is a trapping horizon \cite{hayward}. Moreover, we can try to determine whether $\mathcal{L}_{l}\Theta_{n}<0$ on the horizon which would mean that it is outer trapping horizon. After simple manipulations one arrives at the following formula
\begin{equation}
\mathcal{L}_{l}\Theta_{n}|_{\mathcal{H}}={\textstyle \frac{2}{D-1}\Lambda-\frac{(D-3)(D-2)}{2}\mu R^{1-D}-\frac{1}{D-5}}\frac{\Delta R^{D-5}}{R^{D-3}}
\end{equation}
and we need to prove that
\begin{equation}\label{Lie-derivative}
{\textstyle \frac{2}{D-1}\Lambda R^{D-3}-\frac{(D-3)(D-2)}{2}\mu R^{-2} < \frac{1}{D-5}\Delta R^{D-5}}\ .
\end{equation}
Integrating the last equation over the $D-2$ dimensional compact subspace spanned by coordinates $x^i$ (thus eliminating the right-hand side in (\ref{Lie-derivative})) we get the necessary condition for the horizon being outer
\begin{equation}
\int {\textstyle \frac{2}{D-1}\Lambda R^{D-3}-\frac{(D-3)(D-2)}{2}\mu R^{-2}} < 0\ ,
\end{equation}
which is satisfied for any $\Lambda\leq 0$ and for positive cosmological constant one shall demand
\begin{equation}
R^{D-1}<{\textstyle \frac{(D-1)(D-2)(D-3)}{4}\frac{\mu}{\Lambda}}\ .
\end{equation}

Alternatively, one can consider (\ref{Lie-derivative}) at the maximum of $R$, where $\Delta R^{D-5}<0$. Notice, that for non-negative cosmological constant the right-hand side of (\ref{Lie-derivative}) is strictly increasing function of $R$, so it has maximal value at maximum of $R$. If the horizon is everywhere non-degenerate ($\mathcal{L}_{l}\Theta_{n}\neq 0$) then it is really an outer trapping horizon.

Next, we will consider a gradient of the horizon hypersurface which in our parametrization reduces to
\begin{equation}
\mathbf{N}=\d r-R_{,u}\d u-R_{,i}\d x^{i}\ .
\end{equation}
We can use the sign of its norm to determine the causal character of the horizon. Using the null D-ad (\ref{D-ad}) the corresponding vector can be expressed in the following form
\begin{equation}
\mathbf{N}={\textstyle \frac{1}{2}}(N^{a}N_{a})\mathbf{l}-\mathbf{n}
\end{equation}
and beside being normal to the horizon $\mathcal{H}$ it is also orthogonal to its $u=const.$ $D-2$ dimensional sections $\mathcal{H}_{u}$. One can introduce a second vector orthogonal to these sections
\begin{equation}
\mathbf{Z}={\textstyle \frac{1}{2}}(N^{a}N_{a})\mathbf{l}+\mathbf{n}\ ,
\end{equation}
which satisfies $N^{a}Z_{a}=0$ and therefore is tangent to the horizon $\mathcal{H}$. Then (inspired by \cite{hayward}) $\mathcal{L}_{Z}\Theta_{n}|_{\mathcal{H}}=0$ holds identically which gives the following equation
\begin{equation}\label{null-condition}
{\textstyle \frac{1}{2}}(N^{a}N_{a})\mathcal{L}_{l}\Theta_{n}+\mathcal{L}_{n}\Theta_{n}=0\ .
\end{equation}
If the outer trapping horizon condition is satisfied ($\mathcal{L}_{l}\Theta_{n}<0$) we need to determine the sign of the second term $\mathcal{L}_{n}\Theta_{n}$. One can consider the higher-dimensional generalization of Raychaudhuri equation \cite{Lewandowski} which in the case of nontwisting and nonshearing solution on the horizon where $\Theta_{n}=0$ simplifies to
\begin{equation}\label{Raychaudhuri}
\mathcal{L}_{n}\Theta_{n}=-R_{ab}n^{a}n^{b}\ .
\end{equation}
Since we are considering only aligned null radiation (in the direction of $\mathbf{l}$ with radiation density $\Phi$) and cosmological constant $\Lambda$ the Ricci tensor in (\ref{Raychaudhuri}) can be written as $R_{ab}=\Lambda g_{ab}+\Phi l_{a}l_{b}$ and substituting into (\ref{Raychaudhuri}) we get
\begin{equation}
\mathcal{L}_{n}\Theta_{n}=-\Phi\ .
\end{equation}
Therefore, for nonnegative radiation density $\Phi$ we conclude that $\mathcal{L}_{n}\Theta_{n}\leq 0$. Since both Lie derivatives in (\ref{null-condition}) are nonpositive it follows that $N^{a}N_{a}\leq 0$ which means that the horizon is either null (for $\Phi=0$) or spacelike (for $\Phi>0$). In the latter case it presents an explicit example of dynamical horizon in higher-dimensional spacetime.

\section{Conclusion and Final Remarks}
We have derived the generalization of the Penrose--Tod equation to higher dimensional Robinson-Trautman spacetimes including cosmological constant and pure radiation. Using several mathematical tools we have proved the existence of its solution for any $\Lambda\leq 0$ and for ${\cal R}>0$. The limitations arising for positive $\Lambda$ (\ref{restriction}) are shown to correspond to similar restriction arising in four-dimensional case that are naturally related to the more complicated horizon structure of relevant spacetimes (e.g. naked singularities). Since the sign of scalar curvature of $(D-2)$-dimensional spatial hypersurface does not restrict their topology as it does in $D=4$ we have included the nonpositive case as well. 

Additionally, we have proved that one can consider this horizon as being a higher-dimensional generalization of trapping and dynamical horizon provided additional conditions are satisfied.

The results show that in terms of the presence of the quasilocal horizons the higher-dimensional generalization shares the same qualitative behavior as the standard four-dimensional Robinson-Trautman spacetime. This provides support for considering the generalization given in \cite{podolsky-ortaggio} to be natural not only mathematically but also physically.

Several important issues were not investigated here, namely uniqueness of the horizon hypersurface and its possible topologies. The question of uniqueness is much harder to solve for $D>4$ because of the nonlinearity in gradient. Due to our parametrization of the horizon the issue of its topology is connected with the topology of the underlying spatial geometry (given by $h_{ij}$) of the $(D-2)$-dimensional manifold $M$. So the obvious starting point should be the classification of Einstein spaces of corresponding dimension. For $D=5$ the $S^{1}\times S^{2}$ (black ring) is ruled out since it cannot be endowed with Einstein metric \cite{Besse} (the second homotopy group has to vanish $\pi_{2}(M)=0$) and the Poincar\'{e} conjecture singles out three-sphere as the only simply connected case. In $D=6$ topological obstructions arise for example due to generalized Gauss-Bonnet theorem relating the Euler characteristic $\chi(M)$ and curvature of compact oriented four-manifold. It turns out that $\chi(M)>0$ and it is zero only in the flat case. This rules out $S^{1}\times S^{3}$. In higher dimensions the restrictions are much weaker (positive Ricci curvature implies finite first homotopy group).

\begin{acknowledgments}
This work was supported by grant GACR 202/09/0772 and the Czech Ministry of Education project Center of Theoretical Astrophysics LC06014.
\end{acknowledgments}

\end{document}